\documentclass[prc,fleqn,twoside,floatfix]{revtex4}
\usepackage{epsfig}
\usepackage{amsmath,amssymb,amsfonts}
\def\Journal#1#2#3#4{{\em #1} {\bf #2}, #3 (#4) }
\def\NPA{{ Nucl. Phys.} A}
\def\NP{{ Nucl. Phys.}}
\def\PRL{Phys. Rev. Lett.}

\def\PRC{{Phys. Rev.} C}

\def\PLB {Phys. Lett. B}
\def\EJA {Eur. Phys. J. A}   
                 
\begin{document}
\title{A Self-Consistent Solution to the Nuclear Many-Body Problem at
Finite Temperature}
\author{T. Frick and H. M\"uther}
\affiliation{Institut f\"ur
Theoretische Physik, \\ Universit\"at T\"ubingen, D-72076 T\"ubingen, Germany}
 
\begin{abstract}
The properties of symmetric nuclear matter are investigated 
within the Green's functions approach. We have
implemented an iterative procedure allowing for a self-consistent 
evaluation of the single-particle and two-particle propagators. 
The in-medium scattering equation is solved for a realistic (non-separable) 
nucleon-nucleon interaction including both particle-particle
and hole-hole propagation. The corresponding 
two-particle propagator is constructed explicitely from the single-particle 
spectral functions. Results are obtained for finite temperatures and an
extrapolation to T=0 is presented.  
\end{abstract}
\pacs{21.65.+f, 21.30.Fe}
\maketitle

\section{Introduction\label{Introduction}}      
The evaluation of the saturation properties of nuclear matter from realistic
models of the nucleon-nucleon (NN) interaction is one of the challenging
testing  grounds for many-body theories of quantum
systems\cite{baldo99,muether00}. The strong short range and tensor components,
which are required in realistic NN interactions to fit the NN scattering data
lead to corresponding correlations in the nuclear wave function. The importance
of these correlations is indicated by the observation that a simple
Hartree-Fock or mean field calculation for nuclear matter at the empirical
saturation density using such realistic NN interactions typically  yields
positive energies rather than the empirical value of -16 MeV per
nucleon\cite{muether00}.

While this argument on the importance of correlation effects in the nuclear
wave function is based on a theoretical calculation only, more empirical
evidence on these short range correlations can be deduced from the analysis of
nucleon knock-out reactions\cite{sick91,batenb}. A recent analysis of the
$(e,e'p)$ reaction on $^{208}$Pb covering a wide range of missing energies
indicates that the occupation numbers for the deeply bound proton states are
depleted by the same amount of about 15 to 20 percent\cite{batenb}. This
depletion of the deeply bound hole states can be identified with the
corresponding depletion of hole  states in nuclear matter with momenta well
below the Fermi momentum\cite{ramos,benhar,knehr}. The spectroscopic factors
for these deep lying hole states should be determined by the tensor and
short-range correlations mentioned above. Long-range correlations, on the other
hand, lead to an additional reduction of the spectroscopic factors for states
close to the Fermi surface\cite{vonderfecht91a,nili96}. Since these long-range
correlations are sensitive to the collective excitation modes of the system,
they should be different in finite nuclei as compared to the infinite system of
nuclear matter.

Various tools have been employed to account for correlations in  the
nuclear many-body wave function. These include the traditional approach, the
Brueckner hole-line expansion\cite{day67}, and variational approaches using
correlated basis functions\cite{akmal,fant98}. Attempts have also been made to
employ the technique of a self-consistent evaluation of Green's
functions\cite{kadanoff,kraeft} to the solution of the nuclear many-body
problem. This method offers various advantages: (i) The single-particle Green's
function contains detailed information about the spectral function, i.e. the
distribution of single-particle strength, to be observed in nucleon knock-out
experiments, as a function of missing energy and momentum. (ii) The method can
be extended to finite temperatures, a feature which is of interest for the
study of the nuclear properties in astrophysical environments. (iii) The
Brueckner Hartree Fock (BHF) approximation, the approximation to the hole-line
expansion which is commonly used, can be considered as a specific approximation
within this scheme.

Attempts have been made to start from the BHF approximation and include the
effects of the hole-hole scattering terms in a perturbative
way\cite{koehler92,koehler93,khalaf}. For a consistent treatment, however, one
should treat the propagation of particle-particle and hole-hole states in the
in-medium scattering equation on the same footing. This turned out to be a
rather ambitious aim.
Starting from a single-particle propagator, which is characterized
for each momentum $k$ by one pole at the quasi-particle energy $\varepsilon(k)$, 
only, the in-medium scattering reduces to the Galitskii-Feynman approach. Trying
to solve this equation
one is confronted with the so-called pairing
instability\cite{ramos,vonderfecht93,alm95,bozek1}.    

These pairing effects can be taken into account by means of the BCS
approach\cite{baldbcs,alm1,elgar}. At the empirical saturation density of
symmetric nuclear matter the solution of the gap equation in the $^3S_1-^3D_1$
partial wave leads to an energy gap of around 10 MeV. Another approach is
to consider an evaluation of the generalized ladder diagrams with ``dressed''
single-particle propagators. This means that the single-particle Green's 
functions are not approximated by one pole term but one tries to account for the
complete spectral distribution. 

Attempts have been made to represent the spectral distribution in terms of three
discrete poles\cite{dewu02} or in terms of four Gaussians\cite{roth,ddnsw}.
Indeed it turns out that such an improved representation of the single-particle
Green's function leads to stable solutions. This finding is supported by the
investigations of Bo\.zek and Czerski\cite{bozek01,bozek02,bozek03}
employing separable interaction models. 

It is interesting to note that the same instabilities also occur in studies of
finite nuclei\cite{heinz}, leading to divergent contributions to the binding
energy from the generalized ring diagrams. These contributions remain finite if
the single-particle propagators are dressed in a self-consistent way.

In the present paper we want to present a method in which the equations for the
one-body and two-body Green's functions for nucleons in nuclear matter are 
solved in a self-consistent way,
keeping track of the complete spectral distribution in the single-particle
Green's function. It turns out that the consideration of finite temperature
helps to stabilize the numerical representation of the spectral distribution.
Therefore we first determine the solution for finite temperature and then
extrapolate to the case of $T=0$.        

After this introduction
we outline the formalism of the evaluation of Green's functions for many-body
systems at finite temperature in Section 2.  The results obtained for nuclear
matter using the charge dependent Bonn potential CDBONN\cite{cdb}
are presented in
Section 3, where we also sketch some of the numerical details. Section 4
contains a short summary and the conclusions.

\section{Green's functions}

In the Green's functions approach, physical quantities are
expanded in terms of single-particle propagators.
In a grand-canonical formulation, the one-particle Green's function
can be defined for both real and imaginary times $t$, 
$t^{\prime}$~\cite{kadanoff}:
\begin{equation}
\label{def_g}
{\mathrm{i}}g({\mathbf{x}},t;{\mathbf{x}}^{\prime},t^{\prime})=
\frac{{\mathrm{tr}}\{\exp[-\beta(H-\mu N)]
{\cal{T}}
[\psi({\mathbf{x}}t) \psi^{\dagger}({\mathbf{x}}^{\prime}t^{\prime})]\}}
{{\mathrm{tr}}\{\exp[-\beta(H-\mu N)]\}},
\end{equation}
where $\cal{T}$ is the time ordering operator. It acts on a product
of Heisenberg field operators
$\psi({\mathbf{x}}t)=e^{{\mathrm{i}}tH}\psi({\mathbf{x}})e^{-{\mathrm{i}}tH}$ 
in such a way that the operator with the largest time argument $t$
(or ${\mathrm{i}}t$ in the case that $t$ is imaginary) is put to the
left. A minus sign is included for each commutation.
The trace is to
be taken over all states of the system with all particle numbers.
$\beta$ is the inverse temperature and $\mu$ is the chemical potential
of the system. Depending whether $t>t^{\prime}$ or $t<t^{\prime}$, the
one-particle Green's function can be expressed by the correlation functions
$g^{>}({\mathbf{x}},t;{\mathbf{x}}^{\prime},t^{\prime})$ or
$g^{<}({\mathbf{x}},t;{\mathbf{x}}^{\prime},t^{\prime})$, respectively,
where the time ordering in eq.~(\ref{def_g}) has been carried out explicitely.
Due to the invariance of the trace under cyclic permutations, it can
be shown that the one-particle Green's functions obeys the following 
quasi-periodicity condition
\begin{equation}
\label{kms}
g({\mathbf{x}},t=0;{\mathbf{x}}^{\prime},t^{\prime})=
-e^{\beta\mu}g({\mathbf{x}},t=-{\mathrm{i}}\beta;{\mathbf{x}}^{\prime},
t^{\prime}).
\end{equation}
A hierarchy of relations defines the equations of motion for the
Green's functions.
The equation of motion for the one-particle Green's
function involves the two-body potential as well as the 
two-particle Green's function
$g_{\mathrm{II}}({\mathbf{x}},t;\cdots;{\mathbf{x}}^{\prime\prime\prime},
t^{\prime\prime\prime})$.
In general, the equation of motion for the $N$-particle propagator will
be coupled to the $(N+1)$-particle propagator, if the Hamiltonian
contains a two-body interaction. A good approximation
scheme for $g({\mathbf{x}},t;{\mathbf{x}}^{\prime},t^{\prime})$ must
be based upon an appropriate truncation for the two-particle propagator.
Introducing the self energy
$\Sigma({\mathbf{x}},t;{\mathbf{x}}^{\prime},t^{\prime})$, the
equation of motion or Dyson equation for the imaginary time one-particle 
Green's function reads
\begin{equation}
\label{eom}
\left[
{\mathrm{i}}\frac{\partial}{\partial{t}}+\frac{\nabla^2}{2m}
\right]
g({\mathbf{x}},t;{\mathbf{x}}^{\prime},t^{\prime})
-\int_0^{-{\mathrm{i}}\beta}
{\mathrm{d}}t^{\prime\prime}\int
{\mathrm{d}}{\mathbf{x}}^{\prime\prime}
\Sigma({\mathbf{x}},t;{\mathbf{x}}^{\prime\prime},t^{\prime\prime})
g({\mathbf{x}}^{\prime\prime},t^{\prime\prime};{\mathbf{x}}^{\prime},t^{\prime})
=\delta({\mathbf{x}}-{\mathbf{x}}^{\prime})
\delta(t-t^{\prime}).
\end{equation}
If the full two-particle propagator is replaced by an
antisymmetric product of one-particle propagators,
$\Sigma({\mathbf{x}},t;{\mathbf{x}}^{\prime},t^{\prime})$ is just 
a Hartree-Fock self-energy $\Sigma^{HF}$, which is real.
In general, the self-energy will contain an additional
complex contribution $\Sigma_c$. We will consider the $T$ matrix
approximation for $g_{\mathrm{II}}$ that contains all
particle-particle and hole-hole ladders.

In a translationally and rotationally invariant system in space and
time, the correlation functions depend only on the difference
variables $|{{\mathbf{x}}-{\mathbf{x}}^{\prime}}|$ and
$t-t^{\prime}$. The real and positive spectral function may be defined
using the Fourier 
transforms of the correlation functions along the real time axis
\begin{equation}
A(k,\omega)=g^>(k,\omega)+g^<(k,\omega).
\end{equation}
Since, in the Fourier space, the condition~(\ref{kms}) can be
transformed to
\begin{equation}
\label{kms_2}
g^>(k,\omega)=e^{\beta(\omega-\mu)}g^<(k,\omega),
\end{equation}
the correlation functions become
\begin{eqnarray}
\label{A_p}
g^>(k,\omega)&=&[1-f(\omega)]A(k,\omega)\\
\label{A_h}
g^<(k,\omega)&=&f(\omega)A(k,\omega),
\end{eqnarray}
where $f(\omega)=[e^{\beta(\omega-\mu)}+1]^{-1}$ is the Fermi function.
The coefficients of the Fourier sum, that
takes into account the quasi-periodicity of 
$g({\mathbf{x}},t;{\mathbf{x}}^{\prime},t^{\prime})$
(cf. eq.~(\ref{kms})), can be expressed using the spectral function 
$A(k,\omega)$
\begin{equation}
\label{spec_g}
g(k,z_{\nu})=\int_{-\infty}^{+\infty}
\frac{{\mathrm{d}}\omega}{2\pi}\, \frac{A(k,\omega)}{z_{\nu}-\omega}.
\end{equation}
$z_{\nu}=\frac{\pi\nu}{-{\mathrm{i}}\beta}+\mu$ are the (fermion) Matsubara
frequencies with odd integers $\nu$. $g(k,z_{\nu})$ can be continued
analytically to all non-real $z$. Using the Plemelj formula, the
spectral function can be written as 
$A(k,\omega)=-2\,{\mathrm{Im}}\,g(k,\omega+{\mathrm{i}}\eta)$, where 
$g(k,\omega+{\mathrm{i}}\eta)$ corresponds to the retarded propagator. 

\begin{figure}
\begin{center}
\epsfig{figure=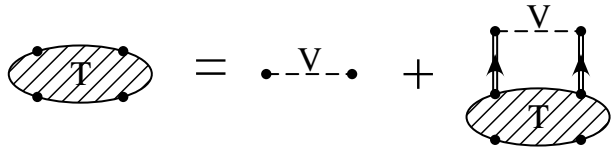,width=8cm}
\end{center}
\caption{\label{t_matrix}The graphical representation of the $T$ matrix}
\end{figure}

By expanding the complex
contribution to the self-energy
in terms of one-particle Green's functions it can be demonstrated 
that it inherits all
analytic properties of $g$. It it thus possible to write
\begin{equation}
\label{spec_Sigma}
\Sigma(k,z)=\Sigma^{HF}(k)-\frac{1}{\pi}\int_{-\infty}^{+\infty}
{\mathrm{d}}\omega\, \frac{{\mathrm{Im}}\Sigma(k,\omega+{\mathrm{i}}\eta)}
{z-\omega}.
\end{equation}
Eq.~(\ref{eom}) is a prescription to determine the Green's function
from the self-energy. In frequency-momentum space, the Dyson equation
is an algebraic equation from which one can derive the spectral
function to be
\begin{equation}
\label{spec_function}
A(k,\omega)=\frac{-2\,{\mathrm{Im}}\,\Sigma(k,\omega+{\mathrm{i}}\eta)}
{[\omega-\frac{k^2}{2m}-{\mathrm{Re}}\,\Sigma(k,\omega)]^2+[{\mathrm{Im}}\,
\Sigma(k,\omega+{\mathrm{i}}\eta)]^2}.
\end{equation}
Because of the $\omega$ dependence of the self-energy, 
the spectral function has not quite a 
Lorentzian shape. The on-shell value of the real and
positive quantity
$\Gamma(k,\omega)=-2{\mathrm{Im}}\,\Sigma(k,\omega+{\mathrm{i}}\eta)$ 
is nevertheless interpreted as the spectral width. 
The next step is to obtain the self energy in terms of the
thermodynamic $T$ matrix. This renormalized interaction takes care of
the correlations induced by the strong short-range and tensor
components of the nuclear two-body force. 
Graphically, the $T$ matrix is depicted in Fig.~\ref{t_matrix}.
Note that the arrows indicate both forward and backward propagating nucleons. 
The analytic structure of the $T$ matrix can be deduced from a product
of two Green's functions, so that
$T^>$ and $T^<$ obey a similar boundary condition as the correlation functions
\begin{equation}
\label{boundary_T}
T^>(\Omega)=e^{\beta(\Omega-2\mu)}T^<(\Omega).
\end{equation}
Like the Green's function, the $T$ matrix can be written in 
a spectral representation
\begin{equation}
\label{spec_T}
T(Z)=V+
\int_{-\infty}^{+\infty}\frac{{\mathrm{d}}\Omega}{2\pi}
\frac{T^>(\Omega)-T^<(\Omega)}{Z-\Omega}.
\end{equation}
Using the Plemelj formula to separate the real and the imaginary part,
one obtains from eqs.~(\ref{boundary_T}) and~(\ref{spec_T})
\begin{eqnarray}
\label{T_bose}
T^>(\Omega)-T^<(\Omega)&=&[e^{\beta(\Omega-2\mu)}-1]\,T^<(\Omega)\nonumber\\
&=&-2\,{\mathrm{Im}}\,T(\Omega+{\mathrm{i}}\eta).
\end{eqnarray}
It is now possible to express ${\mathrm{Im}}\Sigma(k,\omega+{\mathrm{i}}\eta)$ in terms of the
retarded $T$ matrix~\cite{kadanoff,kraeft}
\begin{eqnarray}
\label{im_sigma}
{\mathrm{Im}}\Sigma(k,\omega+{\mathrm{i}}\eta)&=&
-\frac{1}{2}\left[\Sigma^>(k,\omega)+\Sigma^<(k,\omega)\right]\nonumber\\
&=&
-\frac{1}{4}\int \frac{{\mathrm{d}}^3k^{\prime}}{(2\pi)^3}
\int_{-\infty}^{+\infty} \frac{{\mathrm{d}}\omega^{\prime}}{2\pi}
\left[
\left<{\mathbf{kk}}^{\prime}|T^>(\omega+\omega^{\prime})|{\mathbf{kk}}^{\prime}\right>
g^<(k^{\prime},\omega^{\prime})
+
\left<{\mathbf{kk}}^{\prime}|T^<(\omega+\omega^{\prime})|{\mathbf{kk}}^{\prime}\right>
g^>(k^{\prime},\omega^{\prime})
\right]\nonumber\\
&=&
\frac{1}{2}\int \frac{{\mathrm{d}}^3k^{\prime}}{(2\pi)^3}
\int_{-\infty}^{+\infty} \frac{{\mathrm{d}}\omega^{\prime}}{2\pi}
\left<{\mathbf{kk}}^{\prime}|
{\mathrm{Im}}T(\omega+\omega^{\prime}+{\mathrm{i}}\eta)|{\mathbf{kk}}^{\prime}\right>
[f(\omega^{\prime})+b(\omega+\omega^{\prime})]
A(k^{\prime},\omega^{\prime}).
\end{eqnarray}  
In the last line of eq.~(\ref{im_sigma}),
eqs.~(\ref{A_p}), (\ref{A_h}) and~(\ref{T_bose}) have been used.
The $T$ matrix elements are anti-symmetrized.
$b(\Omega)=[e^{\beta(\Omega-2\mu)}-1]^{-1}$ is the Bose
distribution function, which appears because hole-hole scattering diagrams are
treated on the same footing as the particle-particle ladders
in the $T$~matrix approach.   
The pole in the Bose function $b(\Omega)$ at $\Omega=2\mu$ is exactly
canceled by a zero in the $T$ matrix~\cite{alm95,alm1} such that the
integrand remains finite as long as the $T$ matrix does not acquire a
pole. 
Such a pole may occur below a critical temperature $T_C$, a phenomenon
which is often referred to as pairing instability.

A useful assumption, especially for low densities, would be to allow for
forward propagation in the $T$~matrix equation only, 
since the phase space for the holes is very small. 
Then, $T^<=0$, the Bose function in eq.~(\ref{im_sigma})
disappears and so do the complications due to the pole in the
$T$~matrix at low temperature. 
The equations describing this approximations can be cast into the
form of the BHF equations for finite temperature if one makes further
simplifying assumptions for the spectral function $A(k,\omega)$.

The determination of the full $T$ matrix requires the
the knowledge of the product of two one-particle Green's 
functions with equal (imaginary) time arguments, as can be seen in the
graphical representation in Fig.~\ref{t_matrix} 
\begin{equation}
g^0_{\mathrm{II}}(k_1,k_2,t-t^{\prime})=
g(k_1,t-t^{\prime})g(k_2,t-t^{\prime}).
\end{equation}
The one-particle Green's functions can be expressed as Fourier
series. Multiplication by $e^{{\mathrm{i}}Z_{\tau}(t-t^{\prime})}$ 
and integration over the time variable from $0$ to
$-{\mathrm{i}}\beta$ yields
\begin{eqnarray}
\label{matsubara}
g^0_{\mathrm{II}}(k_1,k_2,Z_{\tau})&=&\frac{1}{-{\mathrm{i}}\beta}
\sum_{\nu=-\infty}^{\infty}g(k_1,z_{\nu})g(k_2,Z_{\tau}-z_{\nu}).
\end{eqnarray}
$Z_{\tau}$ is the sum of two fermion frequencies, 
$Z_{\tau}=\frac{\pi\tau}{-{\mathrm{i}}\beta}+2\mu$,
with even integers $\tau$. 
The spectral representation of the Matsubara Green's
functions~(\ref{spec_g}) can be inserted into eq.~(\ref{matsubara})
and the Matsubara sum is converted into a contour integration as described
in Ref.~\cite{kraeft}. The result is
\begin{equation}
\label{two_pp}
g^0_{\mathrm{II}}(k_1,k_2,Z_{\tau})=
\int_{-\infty}^{+\infty}\frac{{\mathrm{d}}\omega}{2\pi}
\int_{-\infty}^{+\infty}\frac{{\mathrm{d}}\omega^{\prime}}{2\pi}
A(k_1,\omega)A(k_2,\omega^{\prime})
\frac{1-f(\omega)-f(\omega^{\prime})}
{Z_{\tau}-\omega-\omega^{\prime}},
\end{equation}
where the relation $f(Z_{\tau}-\omega)=1-f(\omega)$ has been applied.
Expression~(\ref{two_pp}) can be continued analytically.
Substituting 
$\omega^{\prime}=\Omega^{\prime}-\omega$, 
the real and the imaginary part of the retarded propagator
can be separated ($\Omega$ real)
\begin{eqnarray}
\label{g2_re}
g^0_{\mathrm{II}}(k_1,k_2,\Omega+{\mathrm{i}}\eta)
&=&-\frac{\mathcal{P}}{\pi}
\int_{-\infty}^{+\infty}\frac{{\mathrm{d}}\Omega^{\prime}}{2\pi}
\frac{{\mathrm{Im}}\,g^0_{\mathrm{II}}(k_1,k_2,\Omega^{\prime}+{\mathrm{i}}\eta)}
{\Omega-\Omega^{\prime}}+{\mathrm{i}}\,{\mathrm{Im}}\,g^0_{\mathrm{II}}(k_1,k_2,\Omega+{\mathrm{i}}\eta),
\end{eqnarray}
where
\begin{equation}
\label{g2_im}
{\mathrm{Im}}\,g^0_{\mathrm{II}}(k_1,k_2,\Omega+{\mathrm{i}}\eta)=
-\pi
\int_{-\infty}^{+\infty}\frac{{\mathrm{d}}\omega}{2\pi}
A(k_1,\omega)A(k_2,\Omega-\omega)
[1-f(\omega)-f(\Omega-\omega)].
\end{equation}

In a partial wave expansion, the $T$ matrix can now be determined as a
solution of a one-dimensional integral equation
\begin{equation}
\label{T_waves}
\left<q|T^{JST}_{ll^{\prime}}(P,\Omega+{\mathrm{i}}\eta)|q^{\prime}\right>=
\left<q|V^{JST}_{ll^{\prime}}|q^{\prime}\right>
+\sum_{l^{\prime \prime}} \int_0^{\infty} 
{\mathrm{d}} k^{\prime}\,k^{\prime 2}
\left<q|V^{JST}_{ll^{\prime \prime}}|k^{\prime}\right>
g^0_{\mathrm{II}}(P,\Omega+{\mathrm{i}}\eta,k^{\prime})
  \left<k^{\prime}|T^{JST+}_{l^{\prime \prime}l^{\prime}}(P,\Omega+{\mathrm{i}}\eta)|q^{\prime}\right>.
\end{equation}
Note that $g^0_{\mathrm{II}}(k_1,k_2,\Omega+{\mathrm{i}}\eta)$ has to
be expressed in terms of the total momentum $P$ and the relative momentum $q$
of the particle pair, which requires an averaging
over the angle between 
${\mathbf{P}}=\frac{1}{2}({\mathbf{k_1}}+{\mathbf{k_2}})$ and 
${\mathbf{q}}=\frac{1}{2}({\mathbf{k_1}}-{\mathbf{k_2}})$. 
This approximation procedure leads to a decoupling of partial waves with
different angular momentum $J$.
The numerical solution of eq.~(\ref{T_waves}) enables us to use any
nuclear two-body potential given in momentum space. 
The summation of the partial waves,
\begin{equation}
\label{sum_parwaves}
	\left<{\mathbf{kk}}^{\prime}\right|
	{\mathrm{Im}}\,T(\Omega+{\mathrm{i}}\eta)
	\left|{\mathbf{kk}}^{\prime}\right>
	=
	\frac{1}{4\pi}\sum_{(JST)l}(2J+1)(2T+1)
	\left<q({\mathbf{k}},{\mathbf{k}}^{\prime})\right|{\mathrm{Im}}\,
	T^{JST}_{ll}(P({\mathbf{k}},{\mathbf{k}}^{\prime}),\Omega+{\mathrm{i}}\eta)
	\left|q({\mathbf{k}},{\mathbf{k}}^{\prime})\right>,
\end{equation}
yields the $T$ matrix in the form that is needed in
eq.~(\ref{im_sigma}).

Finally, the Hartree-Fock contribution has to be added to the real
part of $\Sigma$
\begin{equation}
\label{hf_sigma}
\Sigma^{HF}(k)
=
\frac{1}{8\pi}\sum_{(JST)l}(2J+1)(2T+1)
\int \frac{{\mathrm{d}}^3k^{\prime}}{(2\pi)^3}
\left<q({\mathbf{k}},{\mathbf{k}}^{\prime})\right|
V^{JST}_{ll}
\left|q({\mathbf{k}},{\mathbf{k}}^{\prime})\right>
n(k^{\prime}),
\end{equation}
where $n(k)$ is the density distribution
\begin{equation}
\label{occupation}
n(k)=
\int_{-\infty}^{+\infty} \frac{{\mathrm{d}}\omega}{2\pi}
f(\omega)
A(k,\omega).
\end{equation}
Note that this is a generalized Hartree-Fock contribution, since the
full one-particle spectral function is applied.

Eqs.~(\ref{T_waves}), (\ref{sum_parwaves}), (\ref{im_sigma}),
(\ref{hf_sigma}), (\ref{occupation}), (\ref{spec_Sigma}), (\ref{spec_function}),
(\ref{g2_im}) and~(\ref{g2_re}) provide a closed system of equations
that have to be solved self-consistently. 

Before the numerical procedure that was applied to solve this system 
is discussed in more detail in the next Section, we will outline two different
approximations to the full approach.

One can think of a simplified set of equations, where the non-trivial spectral
functions are replaced by the quasi-particle
expression 
\begin{equation}
\label{qp_sf}
A(k,\omega)=2\pi\,\delta(\omega-\epsilon(k))
\end{equation}   
both in the two-particle propagator, $g^0_{\mathrm{II}}$, given by
eq.~(\ref{two_pp}) and also in eq.~(\ref{im_sigma}).
This delta type spectral function is peaked at the quasi-particle
energy and introduces an energy momentum relation to the model.
The quasi-particle energy spectrum $\epsilon(k)$ is derived from the
following on-shell condition 
\begin{equation}
\label{qp_energy}
\epsilon(k)=\frac{k^2}{2m}+{\mathrm{Re}}\Sigma(k,\epsilon(k)).
\end{equation} 
We will refer to this scheme as quasi-particle scheme. The reduced
system of equations can be found in Refs.~\cite{schnell,bozek99}. 
Such calculations have been performed e.g. by Alm 
\textit{et al.} for simple separable potentials at finite
temperatures~\cite{alm95}. They find a critical temperature $T_C$,
below which the system undergoes a phase transition to a superfluid state.
Using realistic potentials, however, we found it very difficult to achieve
convergence in the quasi-particle scheme even for $T>T_C$, since for a
wide range of low momenta, eq.~(\ref{qp_energy}) has no unique
solution, which indicates the limitations of the quasi-particle
picture. This problem was also found in Refs.~\cite{vonderfecht93,dewulf}.

The BHF equations at
finite temperature were formulated by Bloch and De
Dominicis~\cite{bloch}.
In this approximation, that was sketched after
eq.~(\ref{im_sigma}), spectral functions
of the quasi-particle type~(\ref{qp_sf}) are applied, too.
With respect to the quasi-particle scheme, the BHF
scheme represents a further simplification, because it does not include
backward propagation in the intermediate states. Since it  
allows for stable solutions, the BHF results are used as a reference
for the full $T$~matrix results.      

\section{Numerical Details and Results}
The self-consistent solution
is obtained in an iterative scheme. 
As a starting point, we use a 
quasi-particle $T$ matrix in the first iteration. 
In the subsequent iteration cycles, 
$g^0_{\mathrm{II}}$ is constructed with the
non-trivial spectral functions given by eq.~(\ref{spec_function}).
We fix the inverse temperature $\beta$ and the chemical potential $\mu$  
and do not change $\mu$ during the
iterative cycle, which implies that the density of the system 
will vary unless self-consistency is achieved. 
The integral equation~(\ref{T_waves}) is solved by a matrix inversion
procedure. However, a pole subtraction as described in Ref.~\cite{haftel} is
unnecessary, because $g^0_{\mathrm{II}}$ has no longer a quasi-particle
form. Nevertheless, the integrand has to be sampled with some care in the
vicinity of the quasi-particle peaks, 
since both the imaginary part and the real part vary rapidly there. 
We use between 40 and 100 integration points for the uncoupled partial
waves.

\begin{figure}
\begin{center}
\begin{minipage}[t]{17cm}
\parbox[t]{8cm}
{
  \epsfig{file=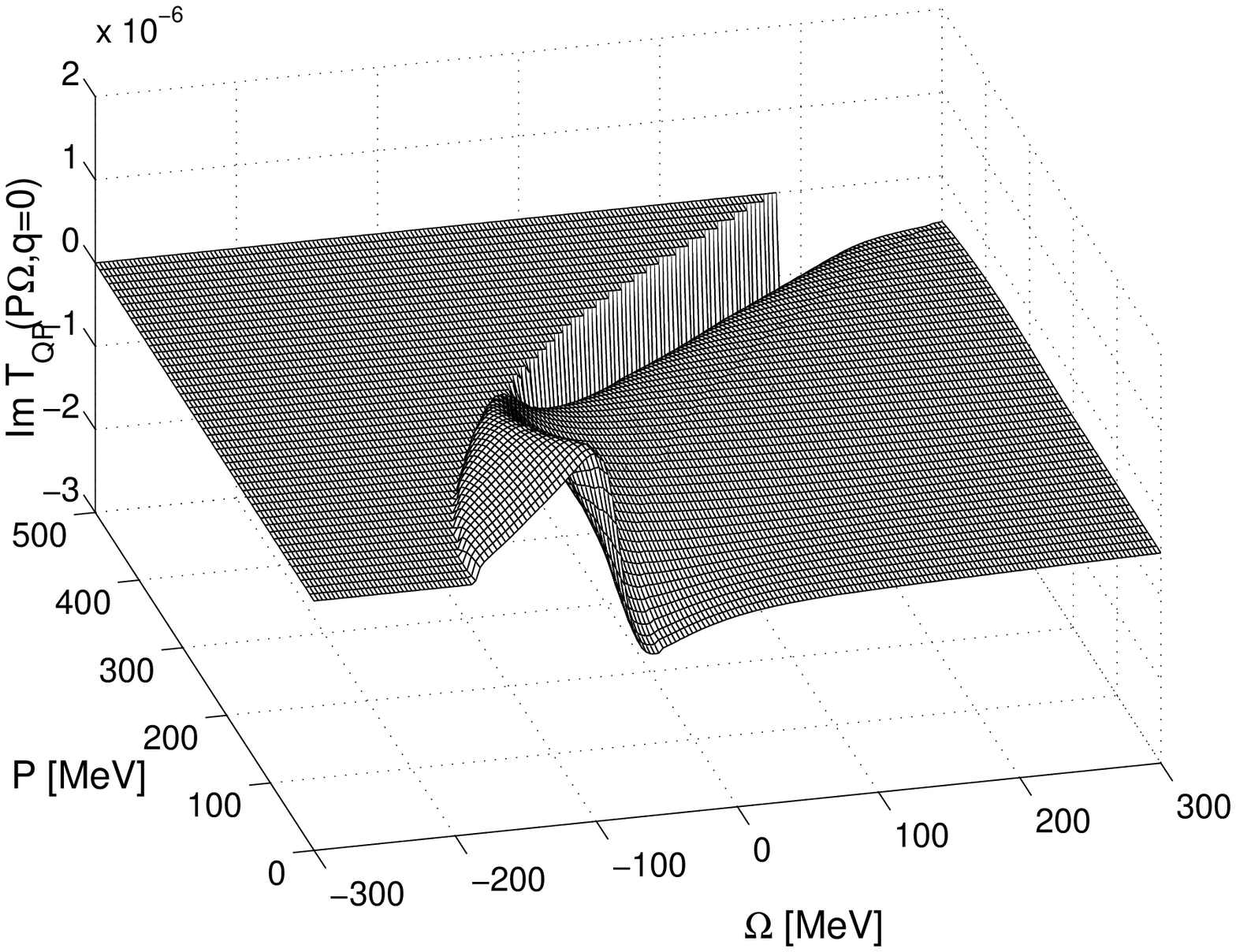,scale=0.4,angle=0}
\begin{center}
Quasi-particle scheme
\end{center}
}
\parbox[t]{8cm}
{
  \epsfig{file=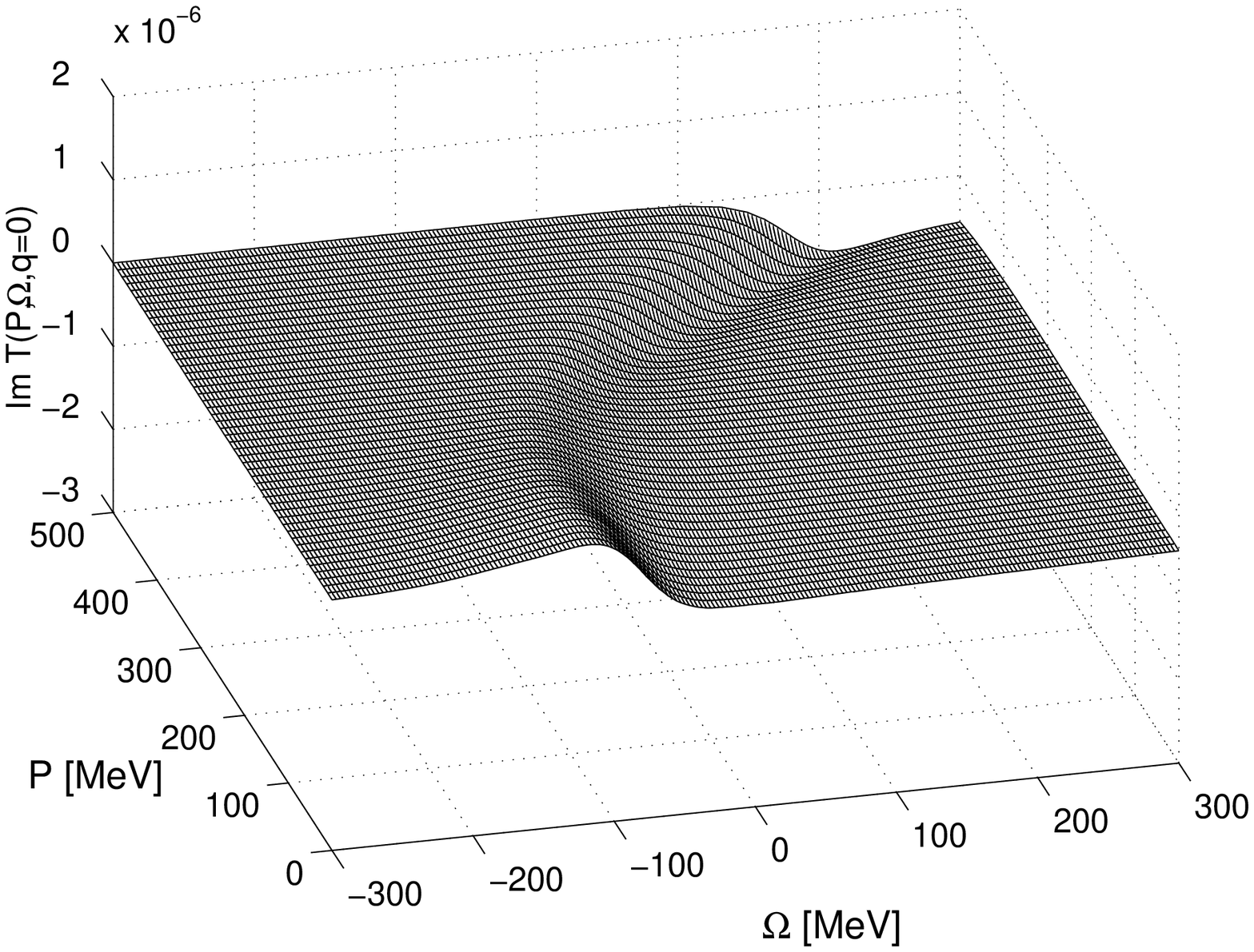,scale=0.4,angle=0}
\begin{center}
Full scheme 
\end{center}
}
\end{minipage}
\end{center}
\caption{\label{ImT}Comparison between the quasi-particle approach
(left panel) and
our full scheme (right panel) on the level of the $T$-matrix. The dependence 
upon the center of mass momentum $P$ and the pair energy $\Omega$ is
displayed. The relative pair momentum $q$ is set to zero. The chemical
potential is $\mu=-40\,\mbox{MeV}$ and the temperature is chosen to be 
$T=10\,\mbox{MeV}$.}
\end{figure}

We have applied the CDBONN potential~\cite{cdb} in our calculations.
All partial waves up to $J=2$ are included in the sum in
eq.~(\ref{sum_parwaves}). Higher partial waves do not contribute
significantly to the off-shell structure of $\Sigma$. 
In contrast, the Hartree-Fock self energy includes
partial waves up to $J=9$.
In Fig.~\ref{ImT}, the thermodynamic $T$~matrix in the
quasi-particle scheme, i.e., using a quasi-particle
$g^0_{\mathrm{II}}$ in eq.~(\ref{T_waves}), is compared to the shape
of the full
$T$~matrix for zero relative momentum of the nucleon pair. 
For pair energies $\Omega<2\epsilon(P)$, the
$T$~matrix has no imaginary part in the quasi-particle approach. 
This becomes obvious if one looks at the quasi-particle approximation
to $g^0_{\mathrm{II}}$,
\begin{equation}
\label{qp_gII}
g^{0}_{\mathrm{II}\,QP}(P,\Omega+{\mathrm{i}}\eta,q)=
\frac{\left<1-f(\epsilon(k_1))-f(\epsilon(k_2))\right>_{\theta}}
{\Omega-\left<\epsilon(k_1)+\epsilon(k_2)\right>_{\theta}+{\mathrm{i}}\eta}.
\end{equation}
An imaginary part in the quasi-particle $T$~matrix can only be formed if
expression~(\ref{qp_gII}) has a pole. For a nucleon pair with zero relative
momentum, the sum of the single-particle energies in the denominator yields
$2\epsilon(P)$. This leads to a sharp structure in the imaginary part of the
$T$~matrix in the quasi-particle approximation for $\Omega\geq 2\epsilon(P)$.
This structure is completely smeared out in the full $T$~matrix
calculation. In this case  ${\mathrm{Im}}T$ is not vanishing also for
 large negative energies values of $\Omega$. 

Once the $T$ matrix is obtained on the $(P,\Omega,q)$-mesh, 
a three-dimensional interpolation has to be applied in order
to carry out the transformation to the integration variables,
($|{\mathbf{k}}^{\prime}|$, $\theta^{\prime}$ and $\omega^{\prime})$, of 
the energy-momentum integrals in eq.~(\ref{im_sigma}).
After the evaluation of the real part of the self-energy with the
principal value integral in eq.~(\ref{spec_Sigma}), 
we interpolate the smooth functions 
${\mathrm{Re}}\Sigma(k,\omega)$ and 
${\mathrm{Im}}\Sigma(k,\omega)$ rather than the spectral function $A(k,\omega)$
to calculate ${\mathrm{Im}}\,g^0_{\mathrm{II}}$.
The careful evaluation of the integral~(\ref{g2_im}) is one of the
crucial steps of the self-consistent procedure. We actually consider 
\begin{equation}
\label{g2_im_act}
{\mathrm{Im}}\,g^0_{\mathrm{II}}(k_1,k_2,\tilde{\Omega}
+{\mathrm{i}}\eta)=
-\pi
\int_{-\infty}^{+\infty}\frac{{\mathrm{d}}\omega}{2\pi}
A(k_1,\omega+\epsilon(k_1))A(k_2,\tilde{\Omega}-\epsilon(k_1)-\omega)
[1-f(\omega+\epsilon(k_1))-f(\tilde{\Omega}-\epsilon(k_1)-\omega)],
\end{equation}   
where $\tilde{\Omega}=\Omega+\epsilon(k_1)+\epsilon(k_2)$.
$\epsilon(k)$ is the on-shell energy, which is defined in the
same way as the quasi-particle energy, eq.~(\ref{qp_energy}).
Note that the peaks of the spectral functions in eq.~(\ref{g2_im_act}) are
located around $\omega=0$ and $\omega=\Omega$, independent of $k_1$ and
$k_2$. This simplifies the construction of the integration mesh.
Additionally, in the subsequent angle-averaging procedure, we can take
advantage of the fact that
${\mathrm{Im}}\,g^0_{\mathrm{II}}(k_1,k_2,\tilde{\Omega})$ is always
peaked around $\Omega=0$. 

A sum rule for the two-particle spectral function 
was given in Ref.~\cite{dickhoff99} for zero temperature.
For the present case, this sum rule can be generalized to 
\begin{equation}
\label{sumrule}
-\frac{1}{\pi}
\int_{-\infty}^{+\infty}\frac{{\mathrm{d}}\Omega}{2\pi}
\,{\mathrm{Im}}\,g^0_{\mathrm{II}}(k_1,k_2,\Omega+{\mathrm{i}}\eta)
=1-n(k_1)-n(k_2).
\end{equation}
This relation was used to check the numerical accuracy that was
achieved for ${\mathrm{Im}}\,g^0_{\mathrm{II}}$ after performing the
integration in eq.~(\ref{g2_im}). Mesh spacings and
integration limits were adjusted such that both sides of
eq.~(\ref{sumrule}) do not deviate by more than $1\%$ 
for single particle momenta up to $k=3000\,\mbox{MeV}$.

After 6-10 iteration cycles, a self-consistent spectral
function $A(k,\omega)$ is obtained. 
We have performed calculations for a range of chemical potentials
between $\mu=-24\,\mbox{MeV}$ and $\mu=5\,\mbox{MeV}$ at
$T=10\,\mbox{MeV}$
and for a range of temperatures between $T=5\,\mbox{MeV}$ and
$T=20\,\mbox{MeV}$ at $\mu=-15\,\mbox{MeV}$.
We found that it is difficult to perform stable and converging
calculations below $T=5\,\mbox{MeV}$ in our scheme.
The reason for these difficulties is that the number of interpolation
and integration mesh points has to be increased strongly in order to obtain
stable results. The spectral functions for lower temperatures exhibit structures
which require a treatment with a larger number of meshpoints.

To extract information for temperatures below $T=5\,\mbox{MeV}$,  we have
extrapolated the retarded self-energy to lower temperatures, using the results
of five stable calculations at $T=5, 7, 10, 15$ and $20\,\mbox{MeV}$. With the 
extrapolated self-energy, we have calculated  the spectral function at
temperatures below $T=5\,\mbox{MeV}$. The reliability of this extrapolation to
lower temperatures will be discussed below.

For the range of densities and temperatures we considered, we have found no
signals of a pairing instability in the full calculation. The signature of
this  pairing instability would be a pole in the $T$~matrix at $\Omega=2\mu$
and zero total momentum of the nucleon pair, $P=0$~\cite{kadanoff}.  Although
the pole appears on the real $\Omega$ axis only at the critical temperature,
$T_C$, the formation of the pole structure can be observed already at higher
temperatures as a precursor effect~\cite{alm95}. Inspecting Fig.~\ref{ImT} we
can see this precursor structure in the quasi-particle $T$~matrix for $P=0$.
This structure is significantly reduced  in the  full
$T$~matrix.   The depletion of the Fermi sea due to short-range correlations,
in addition to the temperature-induced reduction of occupation at the Fermi
surface, weakens the pairing correlations  and no indications for a transition
to superfluidity are found for the explored range of temperatures. Of course,
we cannot exclude such a transition for lower temperatures and/or lower
densities.

\begin{figure}
\begin{center}
\epsfig{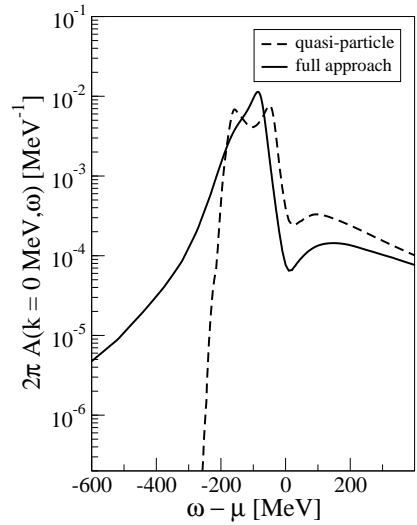}
\end{center}
\caption{\label{cmpqp} Spectral functions for nucleons with momentum $k=0$ at a
temperature $T=10\,\mbox{MeV}$. The spectral function of the self-consistent
calculation (solid line) has been determined for $\mu=-15\,\mbox{MeV}$, which
corresponds to a density of $\rho=0.36\,\mbox{fm}^{-3}$. The results for the
quasi-particle approximation (dashed line) are taken from a calculation at 
$\rho=0.3\,\mbox{fm}^{-3}$.}
\end{figure}

Characteristic differences between the quasi-particle approximation and the full
$T$-matrix approach can also be seen in Fig~\ref{cmpqp}, which shows a
comparison of the spectral  functions for nucleons with momentum $k=0$ in
nuclear matter of a density $\rho$ around 0.3 fm$^{-3}$ at temperature $T$ = 10
MeV. While the quasi-particle spectral function tends to zero below
$\omega-\mu=200\,\mbox{MeV}$, the self-consistent result shows a large tail at
negative energies. This redistribution of strength  is related to the tail that
was found in the imaginary part of the full $T$~matrix in Fig~\ref{ImT}. In the
full calculation one does not constrain e.g.~the construction of the self-energy
at energies below the Fermi energy to
the admixture of two-hole one-particle  configurations, as it is done in the
quasi-particle approach, but the self-consistent evaluation allows for general
n-hole  (n-1)-particle  configurations.   The double-hump structure in the
quasi-particle spectral function reflects the fact that, for low lying states,
eq.~(\ref{qp_energy}) has no unique solution (cf. paragraph after
eq.~(\ref{qp_energy})). The self-consistent spectral functions do not show this
feature any more. The double hump can be interpreted  as an indication for a
strong coupling of the one-hole configuration to  two-hole one-particle
excitations. Obviously this coupling is reduced in the self-consistent
calculation.

\begin{figure}
\begin{center}
\epsfig{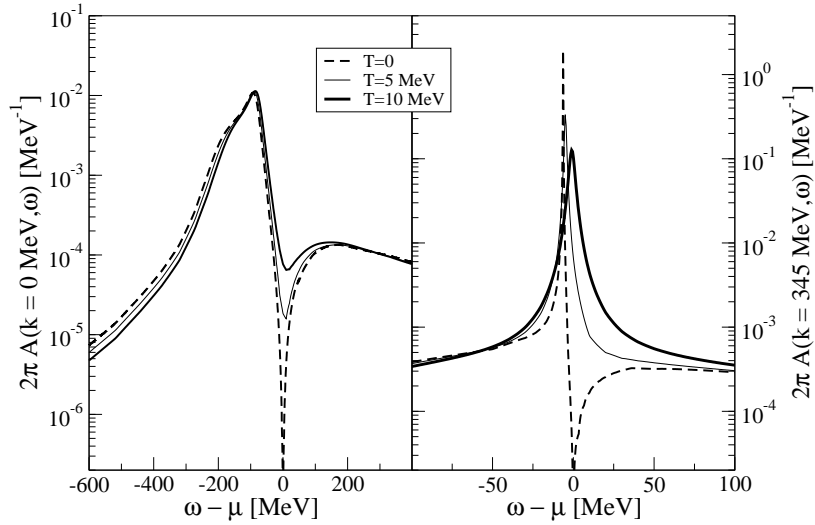}
\end{center}
\caption{\label{SF}Spectral functions in nuclear matter as a function
of the energy at different temperatures. The left panel shows results for
$k=0$, while the spectral functions displayed in the right panel have been
evaluated for $k=345\,\mbox{MeV}$, a momentum just below the Fermi momentum for 
the density $\rho=0.36\,\mbox{fm}^{-3}$, which corresponds to 
$\mu=-15\,\mbox{MeV}$.}
\end{figure}

Spectral functions for nuclear matter at different temperatures are displayed
in Fig.~\ref{SF} considering a value for $\mu=-15\,\mbox{MeV}$, which
corresponds  to a density of about 0.36 fm$^{-3}$. The left panel presents the
spectral function for low-lying hole states with momentum $k=0\,\mbox{MeV}$.
Note that the spectral function for $T=0$, which has been obtained by
extrapolating the self-energy from self-consistent results evaluated for $T$
larger than 5 MeV, shows the correct behavior and vanishes for $\omega$ equal
to $\mu$, separating the hole and particle part of the spectral function. This
separation is of course smeared out at finite temperatures. The width of the
spectral distribution is large for $k=0$ at all temperatures and  almost no
broadening effect due to the temperature can be observed. This can also bee seen
from Table~\ref{table_temp}, which lists results for this width.
 
The spectral function for loosely bound hole states is displayed in the right
panel of Fig.~\ref{SF} assuming a momentum of $k=345\,\mbox{MeV}$, which is just
below the Fermi momentum of the density under consideration. In this case the
width of the spectral distribution is much smaller than for $k=0$. Note the
different energy scale in this part of the figure.  
With increasing $T$, the peak of the spectral function clearly 
broadens and it is shifted to higher energies.
At $T=10\,\mbox{MeV}$, it is almost symmetric around $\omega=\mu$.
This means that at this temperature one does not observe any dip in 
the spectral function at the energy $\omega=\mu$. 

The binding energy per particle, $E/A$ was calculated from the
self-consistent spectral function using the Koltun's sum
rule:
\begin{equation}
\label{eda}
\frac{E}{A}=\frac{d}{\rho}
\int \frac{{\mathrm{d}}^3k}{(2\pi)^3}
\int_{-\infty}^{+\infty} \frac{{\mathrm{d}}\omega}{2\pi}
\frac{1}{2}\left(\frac{k^2}{2m}+\omega\right)A(k,\omega)f(\omega),
\end{equation}
where $d=4$ is the spin-isospin degeneracy factor for symmetric nuclear
matter. A corresponding integral can be used to determine the kinetic energy per
nucleon, $E_{kin}/A$, which implies that we can also evaluate the potential
energy $E_{pot}/A$. 
The density $\rho$ of the system is given by
\begin{equation}
\label{density}
\rho=d
\int \frac{{\mathrm{d}}^3k}{(2\pi)^3}
n(k) = \frac{d}{6\pi^2}\left(k^*\right)^3\,.
\end{equation} 
Both energy and density converged to a high degree of accuracy of
typically better than $0.1\%$ by the end of the iteration.
The temperature dependence of some energy observables is given in
Table~\ref{table_temp}. 
All values at $T=5\,\mbox{MeV}$ and above were obtained from
eq.~(\ref{eda}) using self-consistent spectral functions.
The values at $T=4\,\mbox{MeV}$ and
$T=3\,\mbox{MeV}$ were calculated from the self-consistent results at $T$ larger
and equal to 5 MeV, extrapolating the results for the self-energies. 
The self-energy is a rather
smooth function, but of course the extrapolation introduces uncertainties.
We estimate an error up to about $3\%$ for the interpolated values.
In order to
test this interpolation, we have also evaluated the total energy per nucleon for
those temperatures, for which a self-consistent result exists, by interpolation
from the remaining temperatures. These energies are listed in the column of
Table~\ref{table_temp}, which is labeled $E/A_{\mbox{ext}.}$ 

 At low $T$, we found an almost 
perfect quadratic dependence between the binding energy $E/A$ and the
temperature. From this dependence we extrapolate the total energy per nucleon at
zero temperature to -17.9 MeV .
The decrease of binding with increasing $T$ is mainly, but not only,
due to an increase of the kinetic energy. 

In the last two columns of Table~\ref{table_temp}, we present the on-shell
width $\Gamma(k,\epsilon(k))$ at $k=0$ and at a 
momentum close to the Fermi surface.
While this width is almost independent on the temperature for the low-lying 
states ($k=0$), it increases strongly with $T$ for the states close to the Fermi
surface with a momentum ($k=k^*$, see eq.(\ref{density})).

The density dependence of the spectral width is shown in
Table~\ref{table_width}. The density of the system and the Fermi
momentum $k^*$ of a filled sphere corresponding to that density are also
reported. While the on-shell width for $k=0$ increases with density,
the width at $k=k^*$ decreases. As we have seen already in
Table~\ref{table_temp}, the width for the low-lying states is mainly
determined by the phase space of the occupied states, which does not
depend on the temperature, but on the density. In contrary, at
the Fermi surface, the width is determined by the 
temperature deformation of the Fermi
surface. Since this deformation is more drastic at lower densities,
the width there is larger.

\begin{figure}
\begin{center}
\epsfig{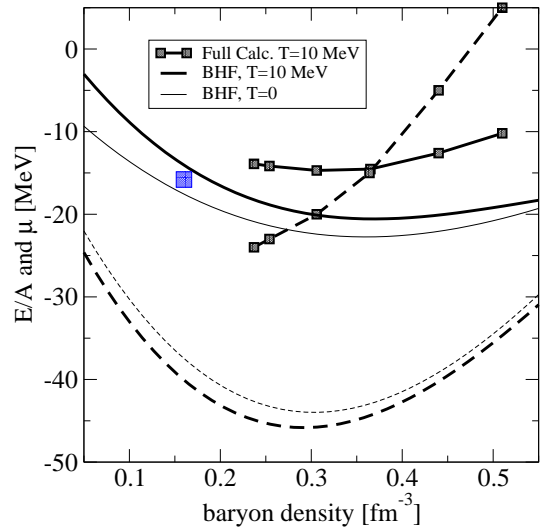}
\end{center}
\caption{\label{EdA}Binding energy per nucleon (solid lines) and the chemical
potential (dashed lines) in symmetric nuclear matter as a function of the
density $\rho$. Results of a full Green's function calculation at a temperature
$T=10\,\mbox{MeV}$ (squares connected by a line), are compared to 
BHF results are shown for zero temperature (thin lines) and
$T=10\,\mbox{MeV}$ (thick lines).
 The large square
represents the `empirical' saturation region that can be 
derived from the Bethe-Weizs\"acker formula.}
\end{figure}

In Fig.~\ref{EdA}, the binding energy per particle $E/A$ (solid line with
squares) and the chemical potential $\mu$ (dashed line with squares) are 
plotted versus the density of the system at $T=10\,\mbox{MeV}$. For a comparison
we also present corresponding results of BHF calculations (continuous choice for
the single-particle spectrum) for 
$T=10\,\mbox{MeV}$ (thick lines) and for zero temperature (thin lines).
We find a repulsive effect in the full $T$~matrix
calculation compared to the BHF calculation at the same
temperature. This repulsive effect increases with density, therefore the
saturation density obtained in the full $T$-matrix calculation is smaller than
the one obtained in the BHF approach. The resulting density, 
$\rho_{sat}=0.31\,\mbox{fm}^{-3}$, however, is still almost 
twice as large as the `empirical value', $\rho_{0}=0.16\,\mbox{fm}^{-3}$.
This repulsive effect relative to the BHF calculations is consistent with the
observations of Ref.~\cite{bozek01}, both with respect to the density
dependence but also with respect to the size of the repulsion.  
 
The so-called Hugenholtz-Van Hove theorem states that, at zero temperature, 
the chemical potential should be equal to the binding energy at the saturation
point~\cite{hugenholtz}. The theorem is
clearly violated by about $20\,\mbox{MeV}$ in the BHF approach, 
while it should be fulfilled in
the self-consistent $T$ matrix approach. This was first shown to be
the case by Bo\.zek and
Czerski for a self-consistent $T$ matrix approach using separable
potentials~\cite{bozek01}. Although a zero temperature
treatment is not really feasible in our approach, we can conclude
a reasonable thermodynamic consistency in our calculation from the fact 
that the intersection point of $\mu(\rho)$ with the binding curve
at $T=10\,\mbox{MeV}$
is located slightly above the saturation density. While the binding energy
will increase as $T$ approaches zero, the chemical potential will become
less attractive for a given density, such that the intersection point
moves closer to the expected region.

\section{Conclusions}  
In this paper, we have presented a method for a self-consistent evaluation of
the one- and two-body Greens function of nuclear matter in the ladder
approximation.  
In contrast to previous works, 
no parameterizations of the single-particle spectral functions were used and 
and the full structure of a realistic NN interaction (CDBONN) was taken into 
account. This requires evaluation of the in-medium NN $T$~matrix, the self-energy
and the spectral functions of the nucleons for a wide range of energies and
momenta. The equations describing these quantities were solved in an
iterative scheme, for a given temperature and chemical potential. 
Self-consistency could be established after several iterations so that
calculation of observables from these Greens functions converged to a high
degree of accuracy. 

This computational scheme works rather well for temperatures above $T$ = 5 MeV. 
The spectral functions for lower temperatures exhibit sharp structures, which
requires a large number of meshpoints for a reliable representation. This
inhibits direct calculations at very low temperatures. Therefore we introduce an
extrapolation procedure, which is based on the smooth behavior of the nucleon
self-energy, to deduce results also for lower temperatures. 

For the range of densities and temperatures we considered,  no
signals of a pairing instability have been observed in the full calculation. The
distribution of strength in the self-consistent spectral function shields the
system against this instability, which is observed in the quasi-particle
approximation. This supports the findings of investigations using separable NN
interactions or parameterizations of the spectral
distribution\cite{ddnsw,bozek01}.
 
Comparing the total energies per nucleon calculated with the self-consistent
$T$-matrix approach with the corresponding results obtained in BHF calculations
we observe a repulsive effect of around 5 MeV per nucleon. As this repulsion
increases with density, the saturation density of the self-consistent Greens
function calculation is shifted to lower densities. The resulting saturation
density, however, is still too large as compared to the empirical value.  

The consistent treatment of hole-hole and particle-particle scattering terms in
the full $T$-matrix calculation leads to results, which are consistent from the
thermodynamic point of view. Therefore, the Hugenholtz-Van Hove theorem, which
is violated in the BHF approach, is respected in the Self-consistent Greens
function approach (see also \cite{bozek01}). 

We would like to acknowledge financial support from the {\it Europ\"aische
Graduiertenkolleg T\"ubingen - Basel} (DFG - SNF).

\begin{table}[t]
\begin{center}
\begin{tabular}{c|cccccc}
\multicolumn{1}{c}{$T$ [MeV]} &
\multicolumn{1}{c}{$E/A$ [MeV]} &
\multicolumn{1}{c}{$E/A_{\mbox{ext}.}$ [MeV]} &
\multicolumn{1}{c}{$E_{kin}/A$ [MeV]} &
\multicolumn{1}{c}{$E_{pot}/A$ [MeV]} &
\multicolumn{1}{c}{$\Gamma(0,\epsilon(0))$ [MeV]} &
\multicolumn{1}{c}{$\Gamma(k^*,\epsilon(k^*))$ [MeV]}
 \\ \hline\hline
20 & -7.05  &   -6.3   & 58.41 & -65.46 & 59.2 & 15.5 \\ 
15 & -11.14 & -11.3 & 54.45 & -65.61 & 58.2 & 9.3 \\
10 & -14.53 & -14.7 & 52.48 & -67.01 & 58.4 & 4.7 \\
7  & -16.32 & -16.2 & 51.87 & -68.17 & 60.5 & 2.5 \\
5  & -17.0  & -17.1 & 50.9  & -67.9  & 62.6 & 1.5 \\
\hline
4 & - & -17.4 & 50.9 & -68.3 & - & - \\
3 & - & -17.6 & 51.0 & -68.6 & - & - \\
\end{tabular}
\caption{\label{table_temp}Energy, $E/A$, kinetic Energy, $E_{kin}/A$, and
potential energy per nucleon, $E_{pot}/A$, calculated for symmetric nuclear
matter for a chemical potential $\mu=-15\,\mbox{MeV}$, which corresponds to a
density of $\rho=0.36\,\mbox{fm}^{-3}$, at different temperatures $T$. The
values at $T=4\,\mbox{MeV}$ and $T=3\,\mbox{MeV}$ were obtained from an
extrapolation of the self-energies evaluated at higher temperatures. Also
displayed are the values for the on-shell width of the self-energy of nucleons
with momentum $k=0$ and $k=k^*=340$ MeV (see eq.(\protect\ref{density})).  
}
\end{center}
\end{table}

\begin{table}[t]
\begin{center}
\begin{tabular}{c|cccc}
\multicolumn{1}{c}{$\mu$ [MeV]} &
\multicolumn{1}{c}{$\rho$ [$\mbox{fm}^{-3}$]} &
\multicolumn{1}{c}{$k^*$ [MeV]} &
\multicolumn{1}{c}{$\Gamma(0,\epsilon(0))$ [MeV]} &
\multicolumn{1}{c}{$\Gamma(k^*,\epsilon(k^*))$ [MeV]} 
 \\ \hline\hline
-23 & 0.254 & 307 & 48.9 & 6.7  \\
-20 & 0.306 & 326 & 53.3 & 5.5  \\
-15 & 0.364 & 345 & 58.4 & 4.7  \\
-5  & 0.44  & 368 & 63.0 & 3.5  \\
5   & 0.51  & 385 & 70.1 & 3.7  \\
\end{tabular}
\caption{\label{table_width}Density-dependence of the on-shell width for
$k=0$ and the momentum $k^*$ (see eq.(\protect\ref{density})). The density
is reported for each chemical potential, too. All calculations were
performed at $T=10\,\mbox{MeV}$.}
\end{center}
\end{table}

\end{document}